\documentclass[conference,compsoc]{IEEEtran}
\IEEEoverridecommandlockouts

\ifCLASSOPTIONcompsoc
  \usepackage[nocompress]{cite}
\else
  \usepackage{cite}
\fi

\ifCLASSINFOpdf
\else
\fi

\usepackage{subfigure}

\ifCLASSOPTIONcompsoc
  \usepackage[nocompress]{cite}
\else
  \usepackage{cite}
\fi
\ifCLASSINFOpdf
  \usepackage[pdftex]{graphicx}
\else
\fi
\ifCLASSOPTIONcompsoc
  \usepackage[caption=false,font=footnotesize,labelfont=sf,textfont=sf]{subfig}
\else
  \usepackage[caption=false,font=footnotesize]{subfig}
\fi
\usepackage[flushleft]{threeparttable}
\usepackage{textcomp}
\usepackage{xcolor}
\usepackage{tabularx}
\usepackage{stfloats}

\usepackage{url}
\usepackage{balance}
\usepackage{flushend}

\usepackage{tablefootnote}
\usepackage{adjustbox}
\usepackage{multicol}

\usepackage[bottom,para]{footmisc}

\usepackage{eso-pic}
\newcommand\AtPageUpperMyright[1]{\AtPageUpperLeft{%
 \put(\LenToUnit{0.5\paperwidth},\LenToUnit{-1cm}){%
     \parbox{0.5\textwidth}{\raggedleft\fontsize{9}{11}\selectfont #1}}%
 }}%
\newcommand{\conf}[1]{%
\AddToShipoutPictureBG*{%
\AtPageUpperMyright{#1}
}
}

\begin{document}
\IEEEpubid{\makebox[\columnwidth]{PRE-PRINT \copyright 2017 IEEE \hfill} \hspace{\columnsep}\makebox[\columnwidth]{ }}

\title{Diluting the Scalability Boundaries: Exploring the Use of Disaggregated Architectures for High-Level Network Data Analysis}

\author{\IEEEauthorblockN{Carlos Vega\IEEEauthorrefmark{1}\IEEEauthorrefmark{2}, Jose Fernando Zazo\IEEEauthorrefmark{1}\IEEEauthorrefmark{2}, Hugo Meyer\IEEEauthorrefmark{3}, Ferad Zyulkyarov\IEEEauthorrefmark{3}, S. Lopez{-}Buedo\IEEEauthorrefmark{1}\IEEEauthorrefmark{2} and Javier Aracil\IEEEauthorrefmark{1}\IEEEauthorrefmark{2}}
\\ \IEEEauthorblockA{\IEEEauthorrefmark{1}Naudit HPCN, \\
Parque Cient\'ifico de Madrid, C/Faraday, 7. 28049 Madrid, Spain \\ Email: \{carlos.vega, josefernando.zazo, sergio, javier.aracil\}@naudit.es}
\\ \IEEEauthorblockA{\IEEEauthorrefmark{2}Escuela Polit\'ecnica Superior, Universidad Aut\'onoma de Madrid, \\ Francisco Tom\'as y Valiente, 11, Madrid, Spain\\
Email: \{sergio.lopez-buedo, javier.aracil\}@uam.es \\
Email: \{carlosgonzalo.vega, josefernando.zazo\}@estudiante.uam.es} 
\\ \IEEEauthorblockA{\IEEEauthorrefmark{3}Barcelona Supercomputing Center \\ Carrer de Jordi Girona, 29-31 08034, Barcelona, Spain \\ 
Email: \{hugo.meyer, ferad.zyulkyarov\}@bsc.es}
}

\maketitle

\conf{IEEE HPCC 2017, Bangkok, Thailand 18 - 20 December, 2017}

\begin{abstract}
Traditional data centers are designed with a rigid architecture of fit-for-purpose servers that provision resources beyond the average workload in order to deal with occasional peaks of data. Heterogeneous data centers are pushing towards more cost-efficient architectures with better resource provisioning. In this paper we study the feasibility of using disaggregated architectures for intensive data applications, in contrast to the monolithic approach of server-oriented architectures. Particularly, we have tested a proactive network analysis system in which the workload demands are highly variable. In the context of the dReDBox disaggregated architecture, the results show that the overhead caused by using remote memory resources is significant, between 66\% and 80\%, but we have also observed that the memory usage is one order of magnitude higher for the stress case with respect to average workloads. Therefore, dimensioning memory for the worst case in conventional systems will result in a notable waste of resources. Finally, we found that, for the selected use case, parallelism is limited by memory. Therefore, using a disaggregated architecture will allow for increased parallelism, which, at the same time, will mitigate the overhead caused by remote memory. \looseness=-1

\end{abstract}

\IEEEpeerreviewmaketitle

\section{Introduction}	
\label{sec:introduction}

\IEEEPARstart{N}{owadays} large data centers serve multitude of different applications, most often via virtual machines (from now on, VMs) that provide an additional level of abstraction. Traditionally, data center servers have been constructed on the basis of hard monolithic building blocks: Motherboards with a fixed number of processor and memory sockets. These building blocks define the characteristics of the VMs that will be run on the data centers. Applications must adapt to the characteristics of the VMs, and scalability is typically horizontal, achieved by instantiating more VMs.

The question is whether this rigid architecture, based on monolithic building blocks, is going to be able to satisfy the requirements of future data centers. A significant increase both in size of data centers~\cite[Fig. 4]{USAeReport}~\cite[Fig. 2]{cisco_forecast} and in volume of data~\cite[Fig. 5.8]{oecd}~\cite[Fig. 4]{cisco_forecast} is expected in the near future. Additionally, upcoming breakthroughs such as the Internet of Things (IoT)~\cite{IoT} and 100 Gbps networks~\cite[pg. 7]{arista}~\cite{100g} will pose new challenges for future data centers.

Actually, the rigidity imposed by a monolithic building block draws a clear border on how computing, memory, storage and network resources may be expanded during future upgrades of a particular data center architecture. Decisions taken during design phase will then condition the way a system evolves, with a direct impact in terms of lower system resource utilization, costly upgrades, and poor energy efficiency.

Recently, disaggregated architectures have been proposed as an alternative to overcome the rigidity of conventional servers. The benefits of disaggregation have been previously discussed in the literature~\cite{BenefitDis}, either by improving vertical elasticity, proposing separate memory blades that disaggregate memory resources~\cite{DisMem}\cite{Marlin}; or with the study of the network capabilities for the disaggregation of resources~\cite{netSup}\cite{NextArch} in data centers. Optical interconnections have also been addressed in the literature~\cite{OptSurv} for improving both low energy consumption in intra-rack interconnections through optical Top of Rack (ToR) switches and inter-rack communications with new optical switch architectures~\cite{energyOpt}.

Furthermore, the uneven distribution of resources in server-oriented architectures impacts energy consumption, which has been well addressed during the last years \textit{``by reducing power draw during idle periods or when at low utilization''}~\cite[pg. 24]{USAeReport}. Disaggregated architectures contribute to this trend offering a finer-grained control over resource provisioning and utilization. In addition to the misallocation of the spatial resources, highly changing workloads over time (e.g. day and night tasks) produce an unbalanced consumption of the available resources. 

However, the cost effectiveness of disaggregation still remains a case of study~\cite{CostEff}, and is hard to quantify at the current stage. However, we strongly believe that cost savings might become one of its major assets due to the better design and low-power components used in these architectures.

For this purpose, the dReDBox\footnote{\url{http://www.dredbox.eu/}} (Disaggregated Recursive Datacentre-in-a-Box) project took the challenge~\cite{dREDi} of breaking the server boundaries aiming to materialize the concept of disaggregation, benefiting itself from the technological improvements of the interconnection components such as low-latency all-optical switches~\cite{highBW}~\cite{1.3Tbps}. The main idea of the dReDBox architecture is to dilute the base unit of data centers through a core of high-speed, low latency optoelectronic fabric that gathers together physically distant components in terms of bandwidth and latency. dReDBox proposes an adaptable low-power data center architecture, moving from the paradigm of mainboard-as-a-unit to a more flexible, software-defined block-as-a-unit schema.

During the development of the dReDBox project, different use cases were studied~\cite{DRBoxDeliver1} as representatives of the very large class of possible applications that the system would host in production. The next subsection focus on the particular case of data analysis and how disaggregated architectures suit their requirements. 

\subsection{Data Analysis in disaggregated architectures}

Data analytics tools usually show a highly varying demand of both processing and storage resources, which usually requires to squeeze either vertical or horizontal scalability, or even both. For instance, indexing a massive amount of documents in NoSQL databases such as Elasticsearch\footnote{\url{https://elastic.co}} or Apache Solr\footnote{\url{http://lucene.apache.org/solr/}} may require heavy memory usage for caching and queuing documents, while, on the contrary, aggregation operations (e.g. calculating interval percentiles) in these search platforms cause a high CPU demand. Thus, resource fragmentation arises under heterogeneous and varying workloads like the aforementioned.

\begin{figure}[!t]
\vspace{-0.5em}
\centering
\includegraphics[width=0.95\columnwidth]{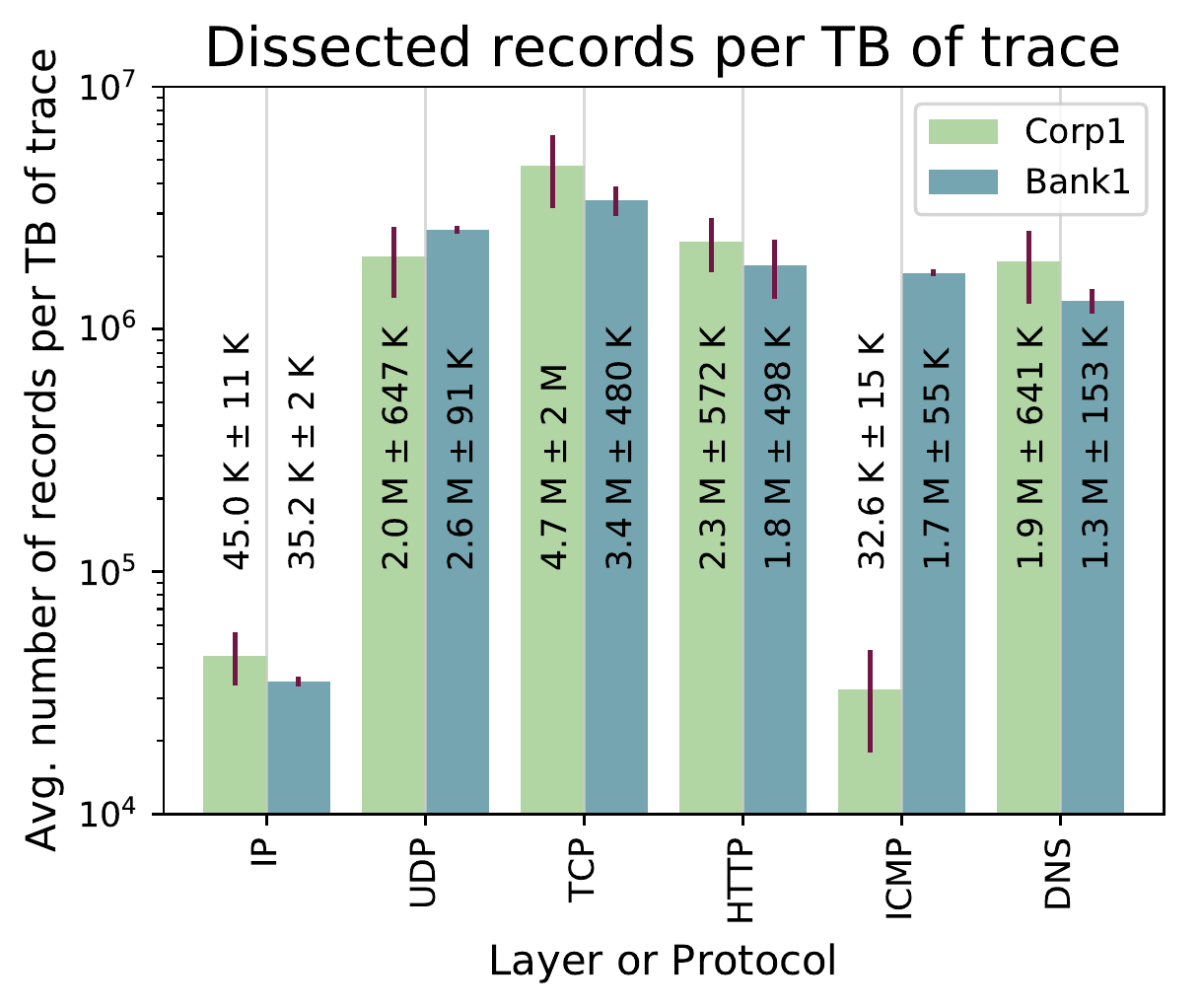}
\vspace{-1em}
\caption{High variation in the number of traffic records per TB of trace from different enterprise networks, obtained by different traffic dissectors.}
\label{fig:perTB}
\vspace{-1em}
\end{figure}

Network data analysis is no exception to the high variability of resource demands. Actually, network traffic in data centers varies widely over time and seasons. Figure~\ref{fig:perTB} shows, for two different data centers, the high variability that exists in the number of records obtained per TB of captured network traffic. In this context, a traffic record corresponds to the series of statistics obtained per flow/transaction/conversation by the traffic dissection tools. There are different dissector tools for each application or protocol layer; in the figure, the number of records from 6 different dissectors (IP, UDP, TCP, HTTP, etc.) is depicted.

Therefore, in traditional data centers, the shape and characteristics of data call for over-dimensioning of systems beyond the actual average needs, in order to deal with occasional workload peaks. This is, provisioning more vertical or horizontal resources than usually needed, if not both of them. Actually, horizontal scaling might not always be simple, as the nature itself of the information poses a challenge for an even distribution of the data. For example, for the network analytics problem, doing an even distribution of Internet flows between several systems is a challenging task. As it is explained in in~\cite{zipf}, \textit{``under certain Zipf-like flow-size distributions, hashing alone is not able to balance workload''}, that is, the straightforward solution of using a hash to horizontally distribute Internet flows is not enough. 

Consequently, the vertical elasticity provided by disaggregated architectures is ideal to alleviate occasional workload peaks in proactive data analysis tasks. For instance, sales in U.S. department stores during the Christmas selling season usually register a 40\% jump from the previous month~\cite{USsales}~\cite{census}. This increment would require higher amount of resources for the analysis of commercial transactions compared with the rest of the year. This seasonal peak of workload may not be worth the costs of having underutilized resources during the rest of the year. However, on a disaggregated data center, workload variations would be lessened, allocating unused remote resources to meet the particular demands.

In this paper we study the feasibility of using disaggregated architectures for intensive data analysis tasks by studying a case of use regarding network analysis. As noted before, disaggregation generally has a throughput penalty regarding remote resource latencies when the ratio of nonlocal operations increases. However, throughput penalties in disaggregated architectures is still an open research problem~\cite{bigdatadis} and is strongly dependent on the task conducted and IO access patterns. 

In light of the above, we now dwell on how the rest of the paper is structured: In Section~\ref{sec:arch} we present the dReDBox disaggregated architecture for data centers. An architecture simulator is then described in Section~\ref{sec:simulator} as the testbed for later evaluations of the use case presented in Section~\ref{sec:usecase}, in which, we will further deepen on the particular data analysis case of high-level network analysis in disaggregated architectures. Finally, we summarize our conclusions about the evaluation and future expectations on the topic.

\section{The dReDBox Architecture}
\label{sec:arch}

To address these disaggregation challenges, the dReDBox project~\cite{dREDi} adopts a vertical architecture, with the lowest level consisting of an interconnection system for remote memory communication that takes full advantage of optical solutions for latencies in the order of tenths of nanoseconds. By interconnecting remote memory controllers and modules with novel scalable optical networks we achieve multi-Tbps-level switch bisection.

Between the many innovation aims of the project, some stand out, particularly:
\begin{itemize}
\item Delivering a novel hyper-visor distributed support to share resources, to allow the execution of commodity VMs, which will be adopted as the execution container. 
\item Making use of a software-defined control for all resources at the hardware programmability level. This hardware orchestration software is to be interfaced via APIs with higher-level resource provisioning, management and scheduling systems.
\item Reducing the power consumption at all layers. The hardware platforms provides an IPMIv2 interface on a per component basis, providing full orchestration control to the hyper-visor. 
\end{itemize}

The dReDBox architecture allows resource usage to flow between the basic unit blocks in the same way the hosting machines provide resource flexibility to processes hosted in VMs, allocating slices of hardware within a server. 

\begin{figure}[!t]
\vspace{-0.5em}
\centering
\includegraphics[width=\columnwidth]{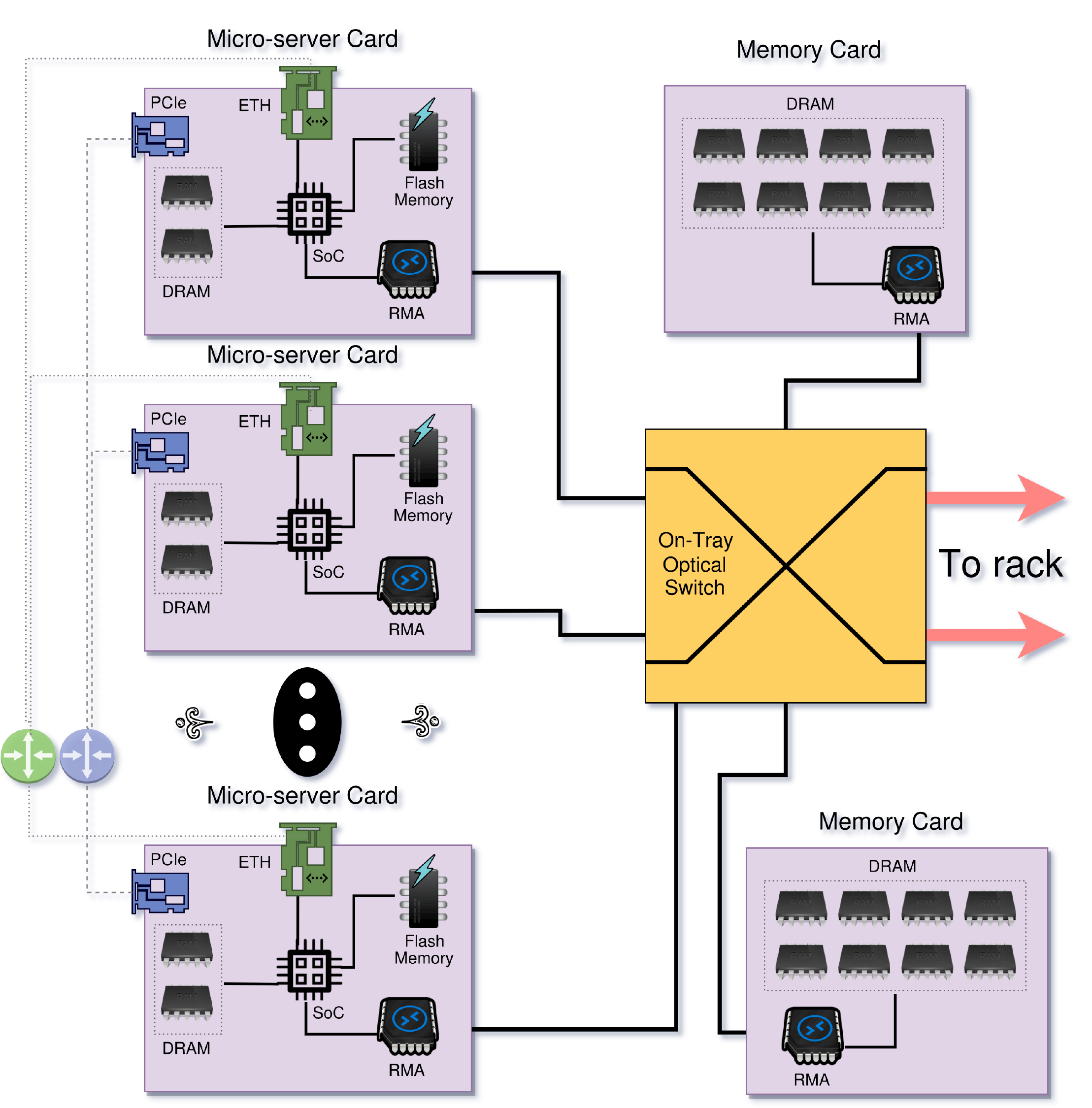}
\caption{Block diagram abstraction of one type of dReDBox server tray. "ETH" stands for Ethernet and "RMA" for Remote Memory Adapters.}
\label{fig:arch}
\vspace{-1em}
\end{figure}

\subsection{Server and Rack Architecture}
The dReDBox platform interfaces different hardware blocks via Remote Memory Adapters (RMA), including a high performance System on a Chip (SoC) for the compute blocks, local memory, flash memory and an ethernet-based Board Management Controller (BMC).

The chosen system architecture for the SoC is based on ARMv8-A Cortex 64-bit~\cite{ARMv8} processors due to its low-power consumption and reduced cost. Although ARM-based application processors can be found in about an 85\% share of mobile devices~\cite{ARMreport} there has been a recent strong interest for the use of ARM cores also in high-performance computing~\cite{HPCARM}~\cite{ARMeuro}. ARM expects that \textit{`25\% share of servers shipped in 2020 will be using ARM-based chips."}~\cite{ARMreport}. 

Figure \ref{fig:arch} depicts a high-level abstraction of the current server architecture, showing how the different components interconnect each other within a generic dReDBox mainboard tray. On the rack-level several trays interconnect themselves through a ToR switch.

\subsection{Electro-optical switched interconnect}

Disaggregation inherently depends on the network performance, which is crucial to serve remote resources. High level of connectivity, bandwidth granularity and low latencies are key for the development of such networks. 

For instance, dense opto-electronic transceiver interfaces have been considered as mid board optics (MBO) due to its low-power consumption and high bandwidth. Manifold variables, such as the transmission band of operation, must be taken into consideration to deliver optimum performance.

With regard to memory interconnection, which always calls for the most demanding latency requirements, dReDBox architecture considers various switching options including all-optical circuit-switching, and also exploring hybrid switching architectures. The number of networked endpoints increases owing to pooling of resources, requiring switch systems at all levels (in tray, cross-rack, in-rack). 

\subsection{Memory Disaggregation}

The dReDBox project disaggregates memory by placing modules on a dedicated memory card and interfacing them over the system and interconnecting them to the remote memory adapter. In this way, dReDBox integrates existing SoC and remote memory components. For that purpose, the development of a memory interface and embodying logic for transmission over the optical network is key. The remote memory component accepts configuration via memory-mapped I/O using special address ranges. Commodity local memory is also available in order to support system bootstrap processes.

\subsection{Operating system support for disaggregation}

The dReDBox platform supports the virtual machine as the execution unit, providing customizable commodity virtual machine execution to applications without compromising their performance. Hence, applications, tools, or systems developed for running in commodity hardware will be able to be deployed without modifications on the disaggregated platform. Aforementioned ARM processors also offer a high level of compatibility to the platform.

The platform hyper-visor is based on \emph{Kernel-based Virtual Machine} (KVM)~\footnote{\url{https://www.linux-kvm.org/}}, a kernel module which enables a standard Linux Operating System to host a number of VMs. Host systems on each dReDBox computing component may not be able to detect all the available components on the platform during local hardware initialization. Actually, they will only have information about the locally attached components. Hence, during bootstrap the host system should retrieve this information from the orchestration tools.

\subsection{Resource allocation and orchestration}

The orchestration in the dReDBox platform is key for an efficient allocation of the data center resources. Forwarding information through the switching network is essential for the interconnectivity of any combination of components. The dReDBox approach provides a physical memory address space available across the whole data center, maintaining a coherent distribution of it, with support for memory ballooning and segmentation. Also, the per component IPMIv2 control offers a potential decrease in the power consumption. A standardized API integrates the orchestration layer with resource management tools. 

Bearing all of this in mind, prior to a virtual machine deployment, a resource scheduling and a platform synthesis step aim to allocate the required resources and set the platform interconnect in an efficient manner.

\section{Simulating a disaggregated architecture}
\label{sec:simulator}

\begin{figure}[b]
	\centering
	\includegraphics[width=\columnwidth]{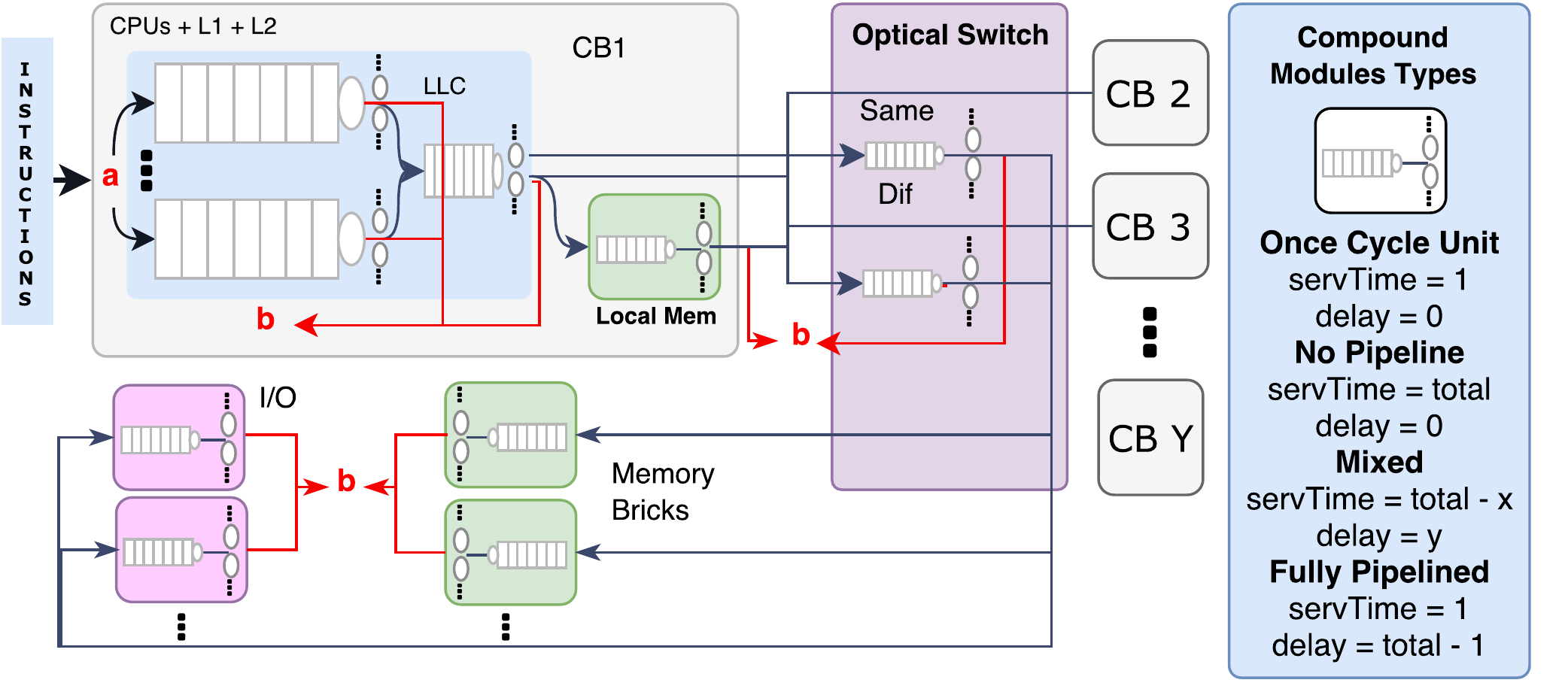}
	\caption{iQ model representing components of the dReDBox platform.}
	\label{fig:iq_model}
\end{figure}

Prior to the discussion of our particular use case, in this section we describe the simulation process used for the evaluation of systems in the dReDBox architecture. This simulator is used to model the behavior of a disaggregated architecture, based on Queue Models. The iQ~\cite{hmeyer2017} model was designed to represent and analyze memory disaggregation, and a statistics-based queuing-based full system simulator was developed to analyze applications performance in disaggregated systems in a quick and accurate manner. These models employ queue structures, message passing, and latency accumulation in order to model such systems. 

Particularly, this process begins when a message is received at the tail of a queue structure. Then, it is propagated until the head of the queue. Finally, a delay is added to emulate the amount of time that the action for that particular message or component takes. For example, in the case of an Integer ALU, the queue models the ALU input queue, where the message passed represents a particular arithmetic instruction (e.g. add, subtract, equals). Consequently, the delay added will depend on the amount of time the ALU unit typically takes to execute that arithmetic instruction. 

Therefore, the simulation process occurs as a collection of discrete events which in our model are then represented as one computational cycle. Execution progress (including instruction generation, propagation, waiting, execution, and retirement as well as resource usage and branch and memory misses) is simulated within the queuing model for every event (i.e every cycle). 

Total performance is measured as a collection of processed messages per total events, i.e., Instructions Per Cycle (IPC). The event driven queuing model we are proposing uses a modular combination of various queue structures, dependency tracking, and probabilistic execution flow to simulate particular systems.

\begin{table}[!t]
\centering
\caption{Memory disaggregation latencies.}
\label{table:sim_values}
\def\arraystretch{1.2}
\begin{tabular}{|c|c|c|}
\hline
\textbf{\begin{tabular}[c]{@{}c@{}}Module \\ Name\end{tabular}}    & \textbf{\begin{tabular}[c]{@{}c@{}}Current \\ Latencies (ns)\end{tabular}} & \textbf{\begin{tabular}[c]{@{}c@{}}Number of accesses \\ per request\end{tabular}} \\ \hline
\begin{tabular}[c]{@{}c@{}}FPGA Addr. Transl.\end{tabular} & 72                                                                         & 1 (CB)                                                                             \\ \hline
Ingress/Egress                                                     & 6.25                                                                       & 4 (2 CB, 2 MB)                                                                     \\ \hline
Network-on-Chip                                                    & 22.4                                                                       & 4 (2 CB, 2 MB)                                                                     \\ \hline
PCS/PMA                                                            & 251                                                                        & 4 (2 CB, 2 MB)                                                                     \\ \hline
DDR4                                                               & 62.5                                                                       & 1 (MB)                                                                             \\ \hline
\end{tabular}
\vspace{-1em}
\end{table}

The queue-model-based methodology emulates processor components by abstracting the implementation details into modular components composed of queue structures, delay parameters and probabilistic driven message generation and event control. 

Figure~\ref{fig:iq_model} shows the bricks' interconnection through an optical switch for simplification purposes. In order to represent processors, memories and other components using queuing models, we have implemented a modular queue structure that models different behaviors through a set of variable configurations (the red \textbf{a} letter in Fig.~\ref{fig:iq_model} indicates the beginning of instruction processing). At the left-bottom side of Figure~\ref{fig:iq_model} the module that is used to represent each component is depicted. This module is formed by a queue, a server and a delay. The queue length and delays required to process instructions are flexible and configurable. The length parameter is used to model resource contention and availability. The service time (\textit{servTime}) represents the time needed to process an instruction until the following instruction may start to be processed. The instruction\textquotesingle s total execution time inside a compound module will be the sum of its own \textit{servTime} plus the service time of all previous executed instructions (pipelining). The lower the service time, the higher level of pipeline and vice versa. The delay (\textit{delay}) parameter is used to complement the \textit{servTime} to ensure the appropriate total delay is added to the instruction. 

Instructions are generated according to the statistical information collected during the profiling stage making use of Linux perf\footnote{\url{https://perf.wiki.kernel.org/}} and pintools such as mica\footnote{\url{http://kejo.be/ELIS/mica/}}. These instructions are then introduced at the entry point of the Compute Bricks as shown in Figure~\ref{fig:iq_model}. Afterwards, depending on the probability values, the instructions will move from one queue to another or to the sink (point \textbf{b} in Fig.~\ref{fig:iq_model}). Instructions move from the different levels of cache and the local memory. In the case that instructions need to access remote bricks (I/O, Memory, etc.), they may need to go through the Optical Circuit Switch (OCS). 

For instance, in~\cite[Table 2]{hmeyer2017} Meyer et al. depict the simulator validation with a comparison between real IPC and simulated IPC for different use cases. These results correspond to past evaluations using previous models but serve as a guide. The IPC is used as a performance indicator in all the experiments, and it helps to analyze the impact of disaggregation in the different use cases.

More specifically, during our evaluation the processor simulated was an Intel\textregistered ~Xeon\textregistered~CPU~E5-2630~0~@~2.30GHz, including all its ALUs and cache levels. The values used to simulate memory disaggregation are presented in Table~\ref{table:sim_values}. This table also includes the number of times that a module is accessed per each remote memory request in the Compute Brick (CB) and in the Memory Brick (MB). The latency values presented in the Current Latencies column of Table~\ref{table:sim_values} are preliminary results which correspond to the current dReDBox prototype currently under development. These latencies as well as the the performance degradation will be reduced during the refinement phase of the project, expecting to halve the total latency.

\section{High-level Network Analysis}
\label{sec:usecase}

\begin{figure}[b]
\centering
\includegraphics[width=\columnwidth]{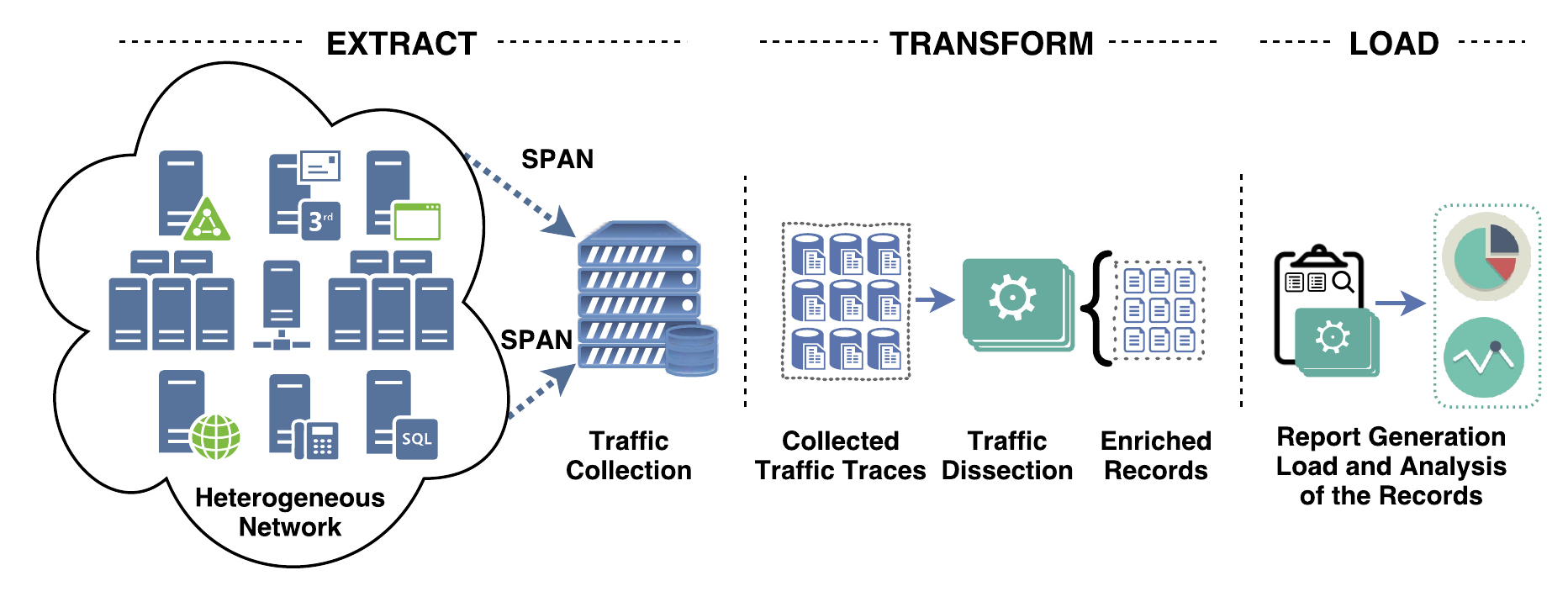}

\caption{Different stages of the high-level network traffic analysis process.}
\label{fig:ETLnet}

\end{figure}

\begin{figure*}[t]
\centering
\begin{adjustbox}{max width=1.25\linewidth,center}
\subfigure[Stress case with high workload.]{\includegraphics[width=1\columnwidth]{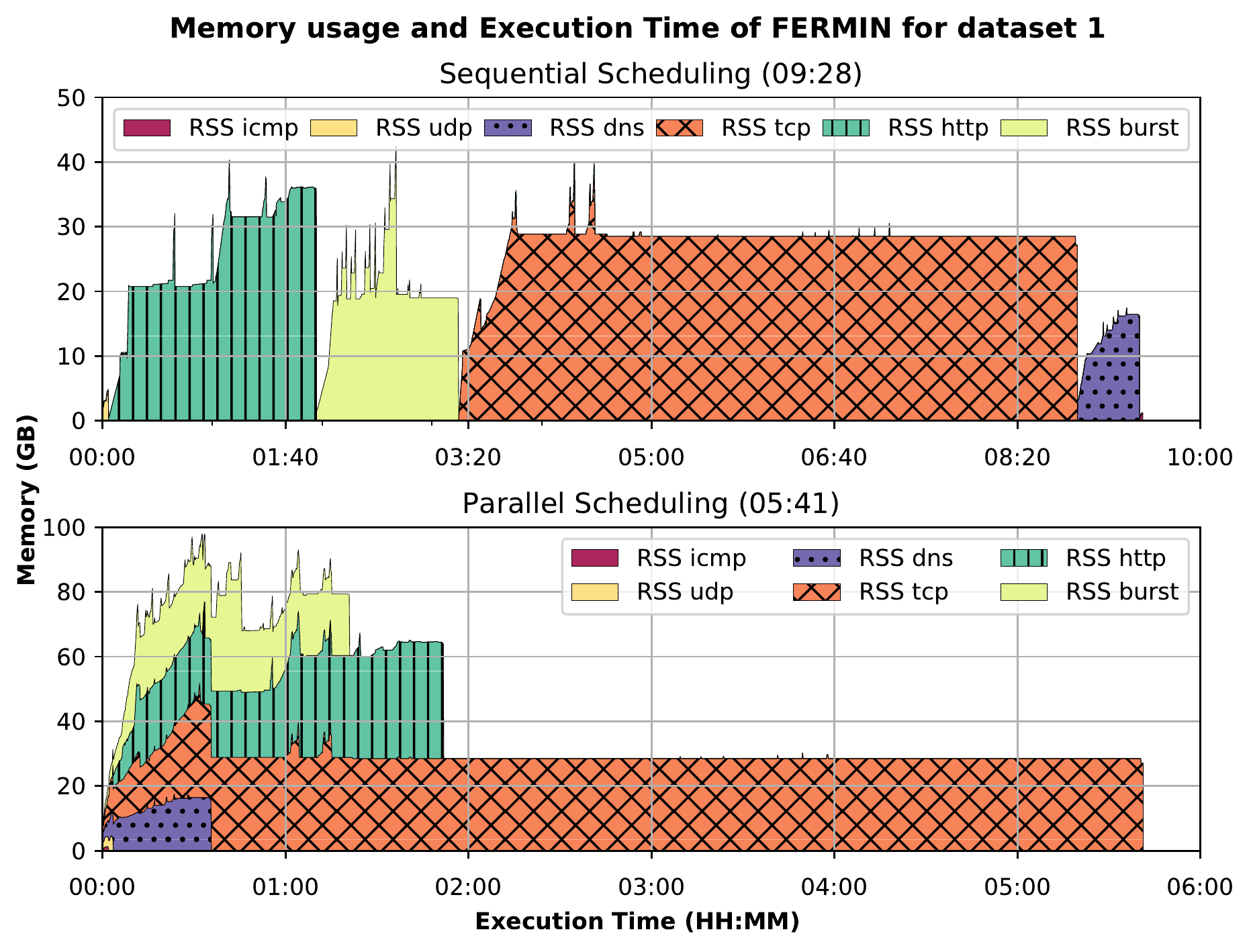}}%
\hfill 
\subfigure[Average workload case.]
{\includegraphics[width=1\columnwidth]{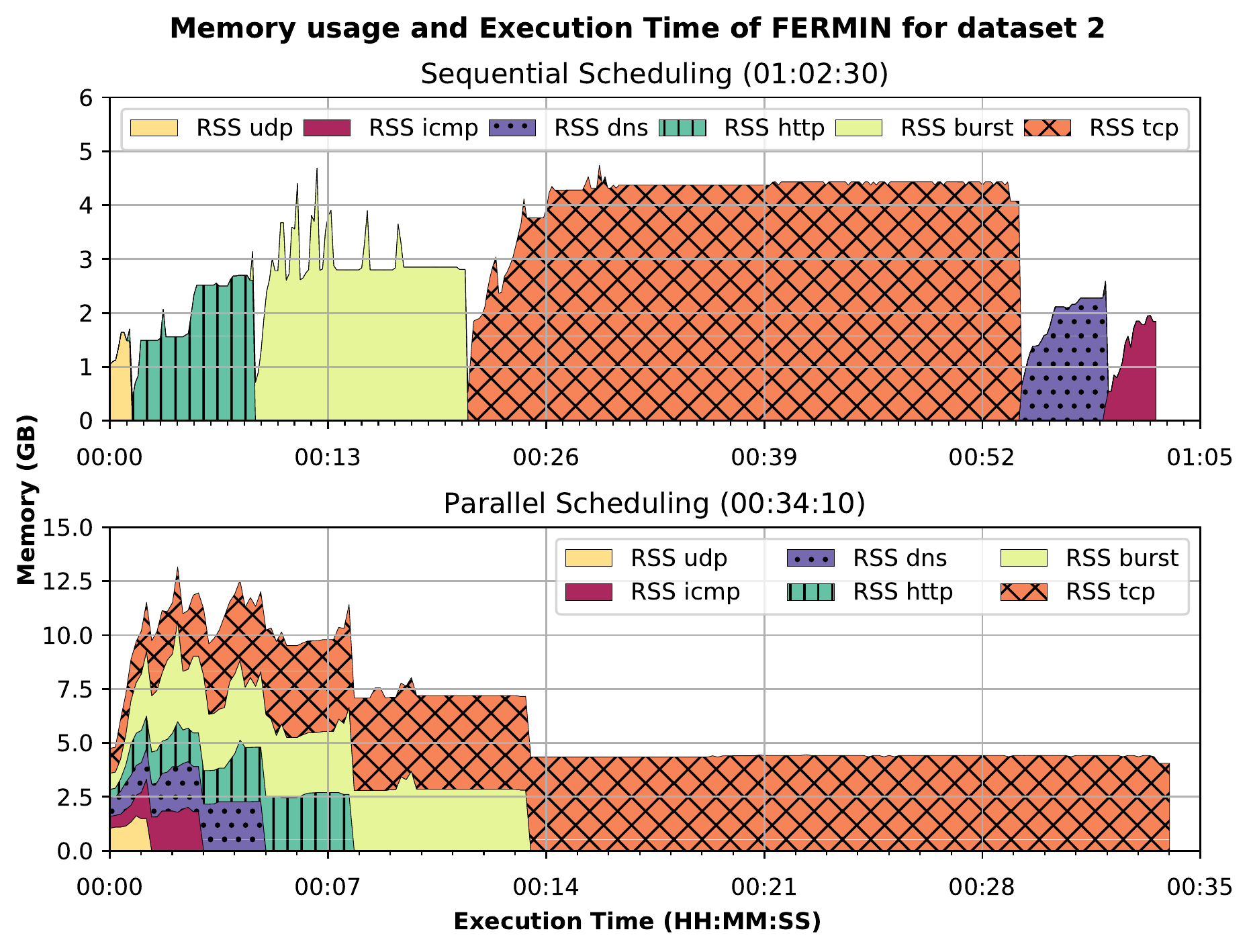}}%
\end{adjustbox}
\caption{Stage-layered execution time and memory usage of the proactive network analysis in different scenarios. RSS stands for Resident Set Size.}
\vspace{-1em}
\label{fig:ferminmem}
\end{figure*}

In order to evaluate the feasibility of using disaggregated architectures for intensive data analysis tasks, a use case of infrastructure analytics is considered. Network monitoring and auditioning is currently a problem very hard to scale when dealing with the current speeds of enterprise backbones (e.g. 10 to 100 Gbps) \cite{HPCAP}~\cite{100g}. 

Nowadays the analytics platform struggles to take full advantage of the motherboard capabilities of conventional systems, because migrating to other boards with more resources has the prohibitive cost of moving data to a different place. Also, due to the nature of the traffic, distribution of the capture and analysis process poses a major challenge. This is why disaggregated and scalable architectures such as the one conceived by the dReDBox project suit the needs of network monitoring and auditioning systems.

For our particular example we considered a traffic analysis solution, \emph{FERMIN}~\footnote{\url{http://www.naudit.es/fermin}} (Factual Executive Report of a Monitored IP Network) for the generation of automated reports aimed to improve the traffic monitoring in large IT infrastructures. 
Needless to say, service outages are one of the main troubles of any data center or network manager. For instance, in a sample of 69 data centers from 43 institutions, the average cost per minute of data center outage was about 7,908~USD~\cite{Ponemon2016}. To anticipate these incidents, proactive traffic analysis is an essential activity in network data center management, helping to proactively identify potential issues and their root cause, before they happen.

In our typical deployment scenario we consider an IT infrastructure that performs IT service continuity management (ITSCM), which conducts a risk analysis for each of the IT services in order to pinpoint the vulnerabilities, assets, threats and countermeasures for each of these services. As part of this process, a traffic analysis is performed either on a daily or weekly basis during the night, by means of an automated process, generating a traffic analysis report. Then, countermeasures and corrective actions are deployed, and monitoring and alarm systems are updated in consequence.

We note that the former are reactive monitoring systems, that in case of an incident, can provide microscopic information about a set of metrics. Usually, they are based on SNMP polling and highlight events like intensive CPU usage or link utilization. Many incidents happen as a consequence of a previously existing problem that was latent and remained undetected. Precisely, such sort of latent and underlying problems are the focus of automatic traffic analysis, which is mostly proactive. Therefore, risk assessment is a long-term proactive activity, based on macroscopic analysis of the traffic and logs. Both approaches are meant to be complementary and serve each other. 

Figure~\ref{fig:ETLnet} depicts this process: Firstly, traffic from the corporate network must be captured through specialized drivers capable of receiving network traffic at 10~Gbps such as the HPCAP network driver~\cite{HPCAP} for Intel\textregistered~Ethernet 10~Gb PCI Express NICs. During traffic dissection, the traffic trace is summarized into enriched records that contain amenable information for the analysis. Traffic dissectors provide records per service (e.g. HTTP, DNS, SQL, etc.), per protocol (e.g. TCP, UDP, ICMP, etc.). The origin of each type of these dissected enriched records may have different nature, since some dissection tools are able to dissect the traffic with streaming mechanisms, off-line techniques, or make use of log records from different services to provide aggregated information. In this respect, there is a clear compromise between real-time delivery of the records and accuracy of the obtained statistics. 

For instance, traffic monitoring probes could provide enriched records in real-time with statistics such as the number of TCP zero window announcements from a given client or server in a particular flow. Nevertheless, not only the zero window announcement events matter, but also how long the client or server was not opening the window. If an endpoint announces a zero window just once in a TCP connection, then it is negligible. Nevertheless, if such an announcement is held for 10~seconds, then chances are that the server is heavily overloaded. As noted before, the latter number of zero windows announcements can be delivered in real time, by means of a counter per flow. However, the blocked time should be obtained off-line, not to increase the probe processing requirements while capturing and storing traffic at high speed.

Is not the purpose of this paper to discuss the particular insights of the tool, but we find necessary to summarize some of its main features. Besides macroscopic traffic analysis, \emph{FERMIN} provides a proactive anomaly detection system to feed back the IT manager with a relevant briefing of the most abnormal and interesting events, all of that through a series of Key Performance Indicators (KPI), such as burst analysis, Red Amber Green (RAG) analysis of different protocols for the main servers, or Topology analysis of both MAC and IP levels, among others. 

In order to address the multiple analysis requirements, which usually demand high-level statistical functions, \emph{FERMIN} was developed in \emph{Python} due to its high versatility and portability, and making use of statistical libraries such as \emph{Numpy}\footnote{http://www.numpy.org/} and \emph{Pandas}\footnote{http://pandas.pydata.org/}, which allow us to update and evolve our system faster. We also developed several modules making use of C language for filtering and helping caching the data and deferring read operations until needed, efficiently.

\begin{figure}[b]
\centering
\includegraphics[width=\columnwidth]{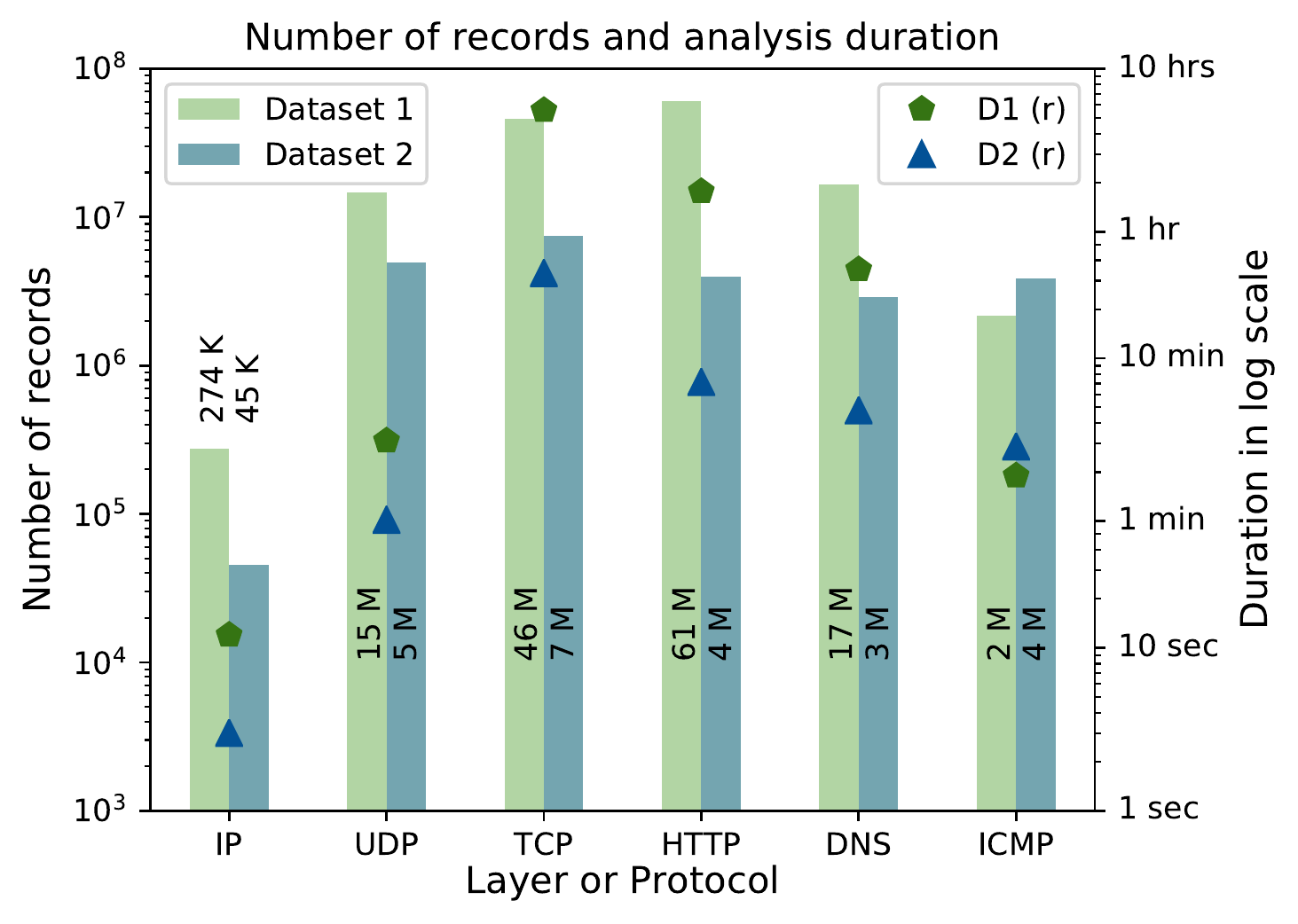}
\caption{Amount of dissected records per layer/protocol together with their corresponding analysis processing time with FERMIN.}
\label{fig:ferminRecTime}
\end{figure}

Although networks show clear day-night or work-holiday behaviors, there are unexpected events that could significantly alter traffic, hence, the computational load of a network analytics problem may be unpredictable. 

For example, Figure~\ref{fig:ferminmem} show two different workload cases of a proactive network analysis conducted in a traditional network probe using an Intel\textregistered~Xeon~E5-2640~v3~@~2.6Ghz with 128GB of RAM. We collected two groups of real network traces from different corporations captured by traffic probes. Both of these groups of traces, \textbf{Dataset~1} and \textbf{Dataset~2}, correspond to Spanish multinational companies. These two groups have different characteristics regarding the traffic. The 6.9 TB traffic traces from the stress case \textbf{Dataset~1} were captured from two different interfaces in the edge of their network. On the other hand, \textbf{Dataset~2} was captured in the the core distribution layer, providing 2 TB of traffic traces, which represent the average amount of traffic during an usual day. 

As observed, the resources needed to address the stress case reach even one order of magnitude more than during the average use case. In fact, in order to tackle with these unpredictable peaks of traffic, we would need to provision quite more resources than usually needed. During peak dates, the main issue is not just the data volume variation but the huge increments in the number of IP conversations, UDP/TCP flows or HTTP transactions. Figure~\ref{fig:ferminRecTime} represents the humongous increase in the number of records dissected corresponding to IP conversations, UDP/TCP flows and HTTP transactions, etc. between the stress and average cases, as well as the differences in the analysis durations.
On a disaggregated architecture workload peaks would be alleviated making use of unused remote resources.

\section{Evaluation, Discussion, and Future Work}

In this light, we conducted a profile evaluation of \emph{FERMIN} in the way described in Section~\ref{sec:simulator}, with the purpose of collecting statistical information about memory access patterns and dependencies between instructions. The simulation was validated through this information with a comparison between real and simulated Instructions Per Cycle (IPC), yielding an error of 4.5\% over the real profiled IPC. Afterwards the application was simulated on the disaggregated architecture model, see the results in Table~\ref{table:sim_results}. The simulated IPC without disaggregation is 1.37 (\textbf{IPC\textsubscript{sim}}). The table presents different results for memory cards with 1, 2, 4, or 8 communication endpoints, each endpoint provides 16~Gbps. The results show that for the current latencies the overhead is about 80\%, which can be reduced to 66\% if latency is halved (expected in future refinements of dReDBox architecture).  

We note that access and provision of remote memory resources should not be the usual behavior but serve to alleviate stress cases such as those described in previous sections. In this sense, the trade-off between the latencies overhead and resource proportionality may be worthwhile for batch-processing systems, including proactive network analysis among others, with high workload variability. 

Such systems do not require hard real-time constraints and can defer their processing tasks in favor of better resource provisioning and scheduling. In those cases, the benefits of a disaggregated architecture is that data analysis tasks, that otherwise would be impossible due to high memory requirements, can be done at the cost of a 66\% performance overhead.

An additional benefit of provisioning remote memory resources is that parallelism is not limited by lack of memory. 

For example, consider Figure~\ref{fig:ferminRecTime} where the execution time and memory usage of FERMIN is depicted for the stress case presented previously, but two different schedules for processes have been evaluated: parallel and sequential. Memory usage for the parallel schedule is notably higher, because at the beginning of the experiment several processes run in parallel, and each one allocates a significant amount of memory. On the contrary, the sequential schedule features a much lower memory usage (peak usage is 40 GB vs. 100 GB for the parallel schedule), but at the cost of almost doubling execution time. That is, if memory resources are limited, parallelism could be jeopardized, because it is going to be the lack of memory and not the number of cores available in the processors what is going to limit parallelism. Moreover, the overhead caused by using remote memory resources can be mitigated by increased parallelism. 

To conclude with, although disaggregation has a clear throughput penalty when the ratio of nonlocal operations increases, this is, the access of remote resources and the induced latencies during the process, we believe that the strengths of such architectures in terms of scalability and resource balance pose a worthwhile compromise. We look forward to the interesting improvements in the interconnection systems, which will add elasticity to the monolithic building block of traditional data centers. Therefore, we expect the valuable information retrieved from our evaluation to help on the refinement phase of the dReDBox architecture.

\begin{table}[!t]
\centering
\caption{Simulation results}
\label{table:sim_results}
\def\arraystretch{1.2}
\begin{threeparttable}
\begin{tabular}{|c|c|c|c|c|}
\hline
\textbf{\begin{tabular}[c]{@{}c@{}}Number \\ of \\ Endpoints\end{tabular}} & \textbf{\begin{tabular}[c]{@{}c@{}}IPC\textsubscript{disagg} \\ with current \\ latencies\end{tabular}} & \textbf{\begin{tabular}[c]{@{}c@{}}Overhead \\ {[}\%{]}\end{tabular}} & \begin{tabular}[c]{@{}c@{}}IPC\textsubscript{disagg} \\ with future \\ latencies\end{tabular} & \begin{tabular}[c]{@{}c@{}}Overhead\tnote{1} \\ {[}\%{]}\end{tabular} \\ \hline
1  & 0.25        & 81.75  & 0.456    & 66.72     \\ \hline
2  & 0.26        & 81.02  & 0.463    & 66.20     \\ \hline
4  & 0.29        & 78.83  & 0.469    & 65.77     \\ \hline
8  & 0.30        & 78.10  & 0.465    & 66.06     \\ \hline
\end{tabular}
\vspace{1em}
\begin{tablenotes}
\centering
\footnotesize
{
\item[1] Overhead~=~$\frac{IPC_{sim}-IPC_{disagg}}{IPC_{sim}} * 100$
}
\end{tablenotes}
\end{threeparttable}
\vspace{-1em}
\end{table}

\section*{Acknowledgments}
This work has been partially supported by the European Union\textquotesingle s Horizon 2020 research and innovation programme under grant agreement No 687632 (dReDBox Project).


\begin{thebibliography}{1}

\bibitem{USAeReport}
Shehabi, A., Smith, S. J., Horner, N., Azevedo, I., Brown, R., Koomey, J., and Lintner, W. (2016). United States data center energy usage report. Lawrence Berkeley National Laboratory, Berkeley, California. LBNL-1005775 Page, 4. \url{https://eta.lbl.gov/publications/united-states-data-center-energy}

\bibitem{cisco_forecast}
Networking, C. V. (2016). {\em Cisco Global Cloud Index}: Forecast and Methodology, 2015-2020. White paper. \url{http://www.cisco.com/c/dam/en/us/solutions/collateral/service-provider/global-cloud-index-gci/white-paper-c11-738085.pdf}
\newblock [Online; accessed 15-Feb-2017].

\bibitem{oecd}
Pe{\~n}a-L{\'o}pez, Ismael and others: \emph{OECD Internet Economy Outlook 2012}, Chapter 4 (2012) \url{http://dx.doi.org/10.1787/9789264086463-en}

\bibitem{IoT}
Al-Fuqaha, A., Guizani, M., Mohammadi, M., Aledhari, M., and Ayyash, M. (2015). Internet of things: A survey on enabling technologies, protocols, and applications. IEEE Communications Surveys and Tutorials, 17(4), 2347-2376. \url{https://doi.org/10.1109/COMST.2015.2444095}

\bibitem{mckinsey}
Manyika, J., Chui, M., Bughin, J., Dobbs, R., Bisson, P., and Marrs, A. (2013). Disruptive technologies: Advances that will transform life, business, and the global economy (Vol. 180). San Francisco, CA: McKinsey Global Institute. \url{http://library.wur.nl/WebQuery/clc/2079131}

\bibitem{arista}
Arista in Q1 2017 \url{https://s2.q4cdn.com/209832288/files/doc_presentations/2017/q1/2017-Highlights-Q1.pdf}

\bibitem{100g}
Zazo, J. F., Lopez-Buedo, S., Sutter, G., and Aracil, J. (2016, November). Automated synthesis of FPGA-based packet filters for 100 Gbps network monitoring applications. In ReConFigurable Computing and FPGAs (ReConFig), 2016 International Conference on (pp. 1-6). IEEE. \url{https://doi.org/10.1109/ReConFig.2016.7857156}

\bibitem{BenefitDis}
Pagès, A., Serrano, R., Perelló, J., and Spadaro, S. (2017). On the benefits of resource disaggregation for virtual data centre provisioning in optical data centres. Computer Communications, 107, 60-74. \url{https://doi.org/10.1016/j.comcom.2017.03.009}

\bibitem{DisMem}
Lim, K., Chang, J., Mudge, T., Ranganathan, P., Reinhardt, S. K., and Wenisch, T. F. (2009, June). Disaggregated memory for expansion and sharing in blade servers. In ACM SIGARCH Computer Architecture News (Vol. 37, No. 3, pp. 267-278). ACM. \url{https://doi.org/10.1145/1555754.1555789}

\bibitem{Marlin}
Tu, C. C., Lee, C. T., and Chiueh, T. C. (2014, October). Marlin: A memory-based rack area network. In Proceedings of the tenth ACM/IEEE symposium on Architectures for networking and communications systems (pp. 125-136). ACM. \url{https://doi.org/10.1145/2658260.2658262}

\bibitem{netSup}
Han, S., Egi, N., Panda, A., Ratnasamy, S., Shi, G., and Shenker, S. (2013, November). Network support for resource disaggregation in next-generation datacenters. In Proceedings of the Twelfth ACM Workshop on Hot Topics in Networks (p. 10). ACM. \url{https://doi.org/10.1145/2535771.2535778}

\bibitem{NextArch}
Greenberg, A., Lahiri, P., Maltz, D. A., Patel, P., and Sengupta, S. (2008, August). Towards a next generation data center architecture: scalability and commoditization. In Proceedings of the ACM workshop on Programmable routers for extensible services of tomorrow (pp. 57-62). ACM. \url{https://doi.org/10.1145/1397718.1397732}

\bibitem{OptSurv}
Kachris, C., and Tomkos, I. (2012). A survey on optical interconnects for data centers. IEEE Communications Surveys and Tutorials, 14(4), 1021-1036. \url{https://doi.org/10.1109/SURV.2011.122111.00069}

\bibitem{energyOpt}
Fiorani, M., Aleksic, S., Casoni, M., Wosinska, L., and Chen, J. (2014). Energy-efficient elastic optical interconnect architecture for data centers. IEEE Communications Letters, 18(9), 1531-1534. \url{https://doi.org/10.1109/LCOMM.2014.2339322}

\bibitem{CostEff}
Abali, B., Eickemeyer, R. J., Franke, H., Li, C. S., and Taubenblatt, M. A. (2015). Disaggregated and optically interconnected memory: when will it be cost effective?. \url{https://arxiv.org/abs/1503.01416}

\bibitem{highBW}
Hasharoni, K. (2014, July). High BW parallel optical interconnects. In Photonics in Switching (pp. PT4B-1). Optical Society of America. \url{https://doi.org/10.1364/PS.2014.PT4B.1}

\bibitem{1.3Tbps}
Hasharoni, K., Benjamin, S., Geron, A., Stepanov, S., Katz, G., Epstein, I., and Mesh, M. (2014, March). A 1.3 Tb/s parallel optics VCSEL link. In SPIE OPTO (pp. 89910C-89910C). International Society for Optics and Photonics. \url{http://doi.org/10.1117/12.2038073}

\bibitem{DRBoxDeliver1}
dReDBox Deliverable D2.1: Requirements specification and KPIs Document \url{http://www.dredbox.eu/deliverables.html}

\bibitem{HTTPD}
Vega, C., Roquero, P., and Aracil, J. (2017). Multi-Gbps HTTP traffic analysis in commodity hardware based on local knowledge of TCP streams. Computer Networks, 113, 258-268. \url{http://doi.org/10.1016/j.comnet.2017.01.001}

\bibitem{NDPI}
Deri, L., Martinelli, M., Bujlow, T., and Cardigliano, A. (2014, August). ndpi: Open-source high-speed deep packet inspection. In Wireless Communications and Mobile Computing Conference (IWCMC), 2014 International (pp. 617-622). IEEE. \url{http://doi.org/10.1109/IWCMC.2014.6906427}

\bibitem{zipf}
Shi, W., MacGregor, M. H., and Gburzynski, P. (2005). Load balancing for parallel forwarding. IEEE/ACM Transactions on Networking (TON), 13(4), 790-801. \url{https://doi.org/10.1109/TNET.2005.852881}

\bibitem{bigdatadis}
Li, C. S., Franke, H., Parris, C., Abali, B., Kesavan, M., and Chang, V. (2017). Composable architecture for rack scale big data computing. Future Generation Computer Systems, 67, 180-193. \\ \url{https://doi.org/10.1016/j.future.2016.07.014}

\bibitem{USsales}
Sales in U.S. Department Stores during the period 2000-2016 \\
\url{https://www.census.gov/econ/currentdata/dbsearch?program=MRTS&startYear=2000&endYear=2016&categories=4521I&dataType=MPCSM&geoLevel=US&notAdjusted=1&submit=GET+DATA&releaseScheduleId=}

\bibitem{census}
The 2015 Holiday Season. U.S. Census CB15-FF.25 \url{https://www.census.gov/newsroom/facts-for-features/2015/cb15-ff25.html}

\bibitem{dREDi} 
Katrinis, K., Zervas, G., Pnevmatikatos, D., Syrivelis, D., Alexoudi, T., Theodoropoulos, D., and Chen, Q. (2016, June). On interconnecting and orchestrating components in disaggregated data centers: The dReDBox project vision. In Networks and Communications (EuCNC), 2016 European Conference on (pp. 235-239). IEEE. \url{https://doi.org/10.1109/EuCNC.2016.7561039}

\bibitem{ARMv8}
Grisenthwaite, R. (2011). ARMv8 technology preview. IEEE Conference.
\url{https://www.arm.com/files/downloads/ARMv8_Architecture.pdf}

\bibitem{ARMreport}
ARM Strategic Report, 2015 v2. \url{https://www.arm.com/company/investors/-/media/arm-com/company/Legacy%20Financial%20PDFs/ARM_Strategic_Report_2015_v2.pdf}

\bibitem{HPCARM}
Rajovic, N., Carpenter, P. M., Gelado, I., Puzovic, N., Ramirez, A., and Valero, M. (2013, November). Supercomputing with commodity CPUs: Are mobile SoCs ready for HPC?. In Proceedings of the International Conference on High Performance Computing, Networking, Storage and Analysis (p. 40). ACM.
\url{https://doi.org/10.1145/2503210.2503281}

\bibitem{ARMeuro}
Durand, Y., Carpenter, P. M., Adami, S., Bilas, A., Dutoit, D., Farcy, A., and Matus, E. (2014, August). Euroserver: Energy efficient node for european micro-servers. In Digital System Design (DSD), 2014 17th Euromicro Conference on (pp. 206-213). IEEE. \url{https://doi.org/10.1109/DSD.2014.15}

\bibitem{HPCAP}
Victor Moreno, Pedro M. Santiago del R{\'\i}o, Javier Ramos, David Muelas,  Jos{\'e} Luis Garc{\'\i}a-Dorado, Francisco J Gomez-Arribas, and Javier Aracil: \emph{Multi-granular, multi-purpose and multi-Gb/s monitoring on off-the-shelf systems}. International Journal of Network Management (2014) \url{http://doi.org/10.1002/nem.1861}

\bibitem{hmeyer2017}
Hugo Meyer, Jose Carlos Sancho, Josue V. Quiroga, Ferad Zyulkyarov, Damian Roca and Mario Nemirovsky.: \emph{Disaggregated Computing. An Evaluation of Current Trends for Datacentres}. International Conference on Computational Science (ICCS) 2017 \url{http://doi.org/10.1016/j.procs.2017.05.129}

\bibitem{Ponemon2016}
Cost of datacenter Outages, datacenter Performance Benchmark Series, Ponemon Institute, January 2016. \url{http://www.ponemon.org/blog/2016-cost-of-data-center-outages}

\end{thebibliography}
\end{document}